\documentclass[smallextended]{svjour3}
\usepackage{graphicx}
\usepackage{mathptmx}
\usepackage{url}
\usepackage{color}

\usepackage{epsfig}

\newcommand{\cm}{cm$^{-1}$}

\newcommand{\ea}{\textit{et al.}}

\newcommand{\eqref}[1]{(\ref{#1})}

\begin{document}
\title{The Status of Spectroscopic Data for the Exoplanet
Characterisation Missions}

\author{Jonathan Tennyson \and Sergei N. Yurchenko}
\institute{Department of Physics and Astronomy, University College London, London WC1E 6BT, UK}

\date{Received: \today / Accepted: date}

\maketitle
\begin{abstract}

  The status of laboratory spectroscopic data for exoplanet characterisation missions such as EChO  is
  reviewed.  For many molecules (eg H$_2$O, CO, CO$_2$, H$_3^+$,
  O$_2$, O$_3$) the data are already available. For the other species
  work is actively in progress constructing this data. Much of the is
  work is being undertaken by ExoMol project (\url{www.exomol.com}).
  This information can be used to construct  a mission-specific
  spectroscopic database.

\end{abstract}

\section{Introduction}

The EChO (Exoplanet Characterisation Observatory) mission \cite{jt523}
aims to use spectroscopy to probe the atmospheres of a range of
transiting exoplanets. To ensure the success of EChO,
or another mission with similar aims, it is
important to have access to the necessary spectroscopic data which
will provide inputs to radiative transport models used for the atmospheres of exoplanets
such as implemented in the codes  NEMESIS \cite{08IrTeKo.model} and TAU \cite{13HoTeTi.exo},
as well as other models such as those of Madhusudhan and Seager \cite{09MaSexx.model}. Codes
designed to model the atmospheres of brown dwarfs and cool stars,
such as ATLAS \cite{70Kuxxxx.model}, MARCS \cite{08GuEdEr.model}, PHOENIX \cite{99HaBaxx.model},
BT-Settl \cite{07AlAlHo.model} and VSTAR  \cite{12BaKexx.dwarfs}, require similar molecular data.
The combination of a comprehensive spectroscopic database and a physically realistic model are essential for
both interpretting the observational data and studying the evolution of the objects.

For our solar system, the
spectroscopic data for modelling radiative transport in planetary
atmospheres is provided by specialist data bases, such as HITRAN
\cite{jt453,jt557} and GEISA \cite{jt504}, which have been compiled
and refined over many years using the data recorded at or about room
temperature, defined by HITRAN as 296~K.  However those exoplanets that will be the
subject of early characterisation studies are
likely to be considerably hotter than the atmospheres of solar systems
planets.  Elevated temperatures lead to both a huge increase in the number of
molecular transitions that need to be considered and, possibly,
changes in the atmospheric composition. It is therefore necessary to
ensure that all data required for modelling such atmospheres will be in place.

There are spectroscopic databases available for hot molecules. Kurucz
has compiled data sets for models of (cool) stellar atmospheres for
many years \cite{11Kurucz.db}. However data are lacking for many
species in these compilations and are approximate for others.  More
recently the high temperature version of HITRAN, known has HITEMP, has
been updated \cite{jt480}. The HITEMP data can be considered as
comprehensive and reliable, but is only available for five species:
water, CO$_2$, CO, OH and NO. The ExoMol project \cite{jt528} is
currently in progress and has the specific aim of providing
spectroscopic data applicable for a large range for temperatures for
studying exoplanet atmospheres.

\begin{figure}[ht]
\begin{center}
{\leavevmode \epsfxsize=0.8\textwidth \epsfbox{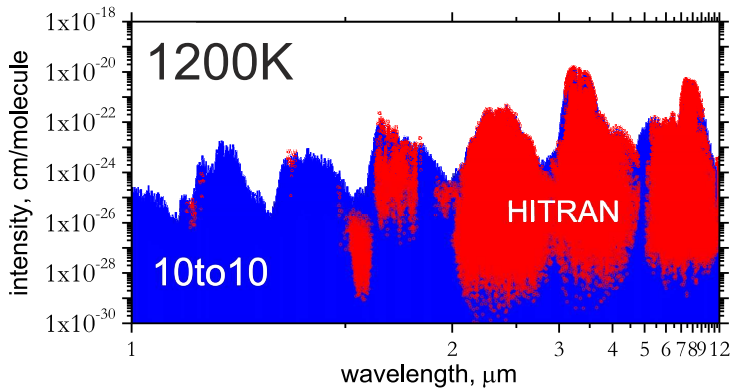}}
\end{center}
\caption{Absorption spectra of CH$_4$  at $T=$ 1200~K simulated using
data from HITRAN 2012 \cite{jt557} and the 10to10 \cite{jt564} line list generated as part of the ExoMol project.  }
\label{f:CH4}
\end{figure}

Hot molecules require significantly more data to simulate spectra at
high temperatures.  Fig.~\ref{f:CH4}  illustrates the importance of the hot
transitions of CH$_4$ missing in HITRAN for modelling the hot
($T=1200$K) absorption. This figure shows an absorption spectrum
simulated using the HITRAN and theoretical 10to10 line lists
\cite{jt564}. The HITRAN spectra, being based on the room temperature
data, suffers from lacking the hot transitions leading to significant
underestimation of the opacity at elevated temperatures, by about
50~\%\ at 1000~K for example.

Table~\ref{molecules} is taken from the initial EChO proposal as submitted in the
autumn of 2010 \cite{EChObid}, see also the 2013 EChO assessment study~\cite{EChOYB}. The
table summarises the main
spectral features that EChO planned to probe. The status of the data
for each of these species is discussed in the next section.
In addition this original compilation~\cite{EChOYB}
we have also identified several potentially important spectral features for the molecules in question,
listed in the last column of Table~\ref{molecules}.
Given the
rapid progress in exoplanetary research and, in particular, the
discovery of new, unanticipated classes of exoplanets, it is possible
that data on further species, not anticipated in the initial proposal
will also be required. The status of some of these data is discussed in
section 3.

\begin{table}[ht]
\caption{ Main spectral features between 0.4 and 16\,$\mu$m. The asterisk indicates  the
molecular/atomic species already detected in the atmospheres of exoplanets.
At wavelengths shorter than 2\,$\mu$m spectroscopic data are often not complete,
so that the use of this region is much more difficult for band identification
and analysis. The main bands are illustrated in bold.}
\label{molecules}
\begin{tabular}{|p{1.5cm}|p{2.5cm}|p{2.5cm}|p{1.5cm}|p{1.5cm}|p{1.5cm}|}
\hline
         & 0.4-1\,$\mu$m & 1-5\,$\mu$m   & 5-11\,$\mu$m   & 11-16\,$\mu$m  & Additional$^a$  \\
\hline
 \emph{R, baseline}       &   300 &   300 &  $\ge$30      &  30  & \\
 \hline
\emph{ R, desired }     &  300  &  300  &  300   &  300  & \\
\hline
\emph{Species}   &        &   &    &    & \\
\hline
C$_{2}$H$_{2}$& -       & 1.52,  \textbf{3.0}            & 7.53  & \textbf{13.7} & 2.5  \\
C$_{2}$H$_{4}$ & -       & \textbf{3.22}, 3.34           & 6.9, \textbf{10.5} & - & \\
C$_{2}$H$_{6}$ & -  & 3.4                               & -      & \textbf{12.1}  & 6.7 \\
CH$_{3}$D       & -     & 3.34, \textbf{4.5}            & 6.8, 7.7, \textbf{8.6} & - & \\
{*}CH$_{4}$     & 0.48,  0.57. 0.6,  0.7, 0.79,  0.86  & 1.65, 2.2, 2.31, 2.37,  \textbf{3.3} &  \textbf{7.7} & - & 6.5 \\
{*}CO   & -     &  1.57, 2.35, \textbf{4.7}      & -     & - & 1.19 \\
{*}CO$_{2}$  &   -       &  1.21, 1.57, 1.6, 2.03,  \textbf{4.25}  & -     & \textbf{15.0} & 1.43, 2.7 \\
FeH     &  0.6-1        &   1-2                         & -     & - & \\
H$_{2}$ & -     & 2.12                          & -     & - & \\
{*}H$_{2}$O     & 0.51, 0.57, 0.65, 0.72, 0.82, 0.94 & 1.13, 1.38, 1.9, \textbf{2.69} & 6.2 &  continuum  & \\
H$_{2}$S        & -     &     \textit{2.5}$^b$, 3.8            &  7.7    & - & 1.3, 1.6, 2.0, 2.6, 4.1 \\
H$_{3}^{+}$ & - & 2.0, 3-4.5                    & -     & - & \\
HCN      & -     & \textbf{3.0}       & -     & \textbf{14.0} & 7.1 \\
HDO     & -     & 2.7,3.67                      & 7.13  & - & \\
N$_{2}$O        & -     &                2.8, 3.9, \textbf{4.5}          & 7.7, 8.5       & - & \\
NH$_{3}$         & 0.55, 0.65, 0.93      & 1.5, 2, 2.25, 2.9, \textbf{3.0}       & 6.1, \textbf{10.5} & - & \\
NO$_{2}$        & -     &                3.4             & \textbf{6.2}, 7.7     &  13.5 & \\
O$_{2}$ &  0.58, 0.69, 0.76, 1.27   &      -                      &  -    &   -  & 1.07, 6.25 \\
O$_{3}$ &  0.45-0.75 (the Chappuis band)        &  4.7                           &  \textit{9.1}$^b$, \textbf{9.6}     & 14.3  & 3.3 \\
PH$_{3}$  & -    & 4.3                           & 8.9, 10.1 & - & \\
SO$_{2}$        & -     &                4               &  \textbf{7.3}, 8.8    & - & \\
TiH     &  0.4-1        &   1-1.6                               & -     & - & \\
TiO     &  0.4-1        &    1-3.5                      & -     & - & \\
VO      &  0.4-1        &    1-2.5                      & -     & - & \\
 Ca &     0.8498,  0.8542,   0.8662      &                       & -     & - & \\
He      & -     & 1.083                         & -     & - & \\
{*}K     & 0.76           & -                             & -     & - & \\
{*}Na    & 0.589 &     1.2                               & -     & - & \\
Rayleigh & 0.4-1 & -            & -     & - & \\
Cloud/haze &   yes      & possible      & silicates, etc. & - & \\
H  H$\alpha$ & \textbf{0.66} & &  &  & \\
H  H$\beta$ & 0.486 & &  &  & \\
\hline
\end{tabular}

$^a$ Additional potentially important strong features.

$^b$ Features from from Ref.~\cite{EChOYB} to be corrected or removed.

\end{table}

\section{Status}

The spectroscopic properties of
the atomic species listed in Table~\ref{molecules}, namely Na, K, H  H$\alpha$, H  H$\beta$,
He, Ca, are all well known and can be found in standard data sources such as NIST
(see \url{http://www.nist.gov/pml/data/asd.cfm}). 

Details of the methodology required for the construction of molecular line lists valid for extended ranges
of
temperatures and wavelengths has been discussed elsewhere \cite{jt528,jt475,jt511} and will not
be repeated here. We also point the reader to recent work on the use high-accuracy, {\it ab initio}
dipole moment surfaces for such studies \cite{13Yuxxxx.method,jt522,jt573} 
Below we consider in turn each molecular
species listed in Table~\ref{molecules}.\\

{\bf C$_{2}$H$_{2}$}: Extensive experimental work on acetylene has been
performed in Brussels, see Moudens \ea\ \cite{11MoGeBe.HCCH} for example, and a line list based on
this work is promised. Initial theoretical studies have been performed in London \cite{jt479}
and a full line list is planned as part of the ExoMol project. At least one line list should
be available in due course.\\

{\bf C$_{2}$H$_{4}$}: there is no hot line list for ethylene. Work on a hot line list for this molecule is in progress as part of the ExoMol
project \cite{jtpav}. A full line list should be available soon.
Experimental (room-temperature) line lists are available for the low frequencies region, below 2400~\cm,
only \cite{10FlLaSa.C2H4,11LaFlTc.C2H4}
 \\

{\bf C$_{2}$H$_{6}$}: there is no hot line list for ethane. One is planned as part of the ExoMol
project and some preliminary calculations have been performed \cite{jtAnatoly}. Ethane has 8 atoms and a low energy
vibrational mode, so the number of lines at elevated temperatures are likely to be
unmanageable.
Experimentally (room-temperature) only the region the infrared active fundamental transitions from the
three main spectral regions, around 12 $\mu$m \cite{07AuMoFl.C2H6}, 6.2 -- 7.5 $\mu$m \cite{08LadiAu.C2H6}, and 3.3 $\mu$m \cite{11LadiVa.C2H6} are covered.
  The proposal will be therefore to provide temperature-dependent cross sections  \cite{jt542}
for ethane; these  should be available in due course.\\

{\bf CH$_{3}$D}: There are no good current line lists for singly deuterated methane and the problem
is harder than methane itself.
A room-temperature line list for the 3 $\mu$m region,
important because it is in the window of the strong CH$_4$ absorption,
is available \cite{06NiChBr.CH3D}.
Work is a more comprehensive CH$_3$D line list, appropriate for higher temperatures,
is planned  both as part of  the ExoMol project
and elsewhere.
A hot line list for  CH$_{3}$D should be available in due course.\\

{\bf CH$_{4}$}: Despite the detection of methane in HD189733b
\cite{08SwVaTi.exo} and elsewhere
\cite{10SwDeGr.exo,11JaCaTh.exo,jt495}, it has been recognised
that the spectroscopic data for methane was inadequate for such studies
\cite{jt198,07ShBuxx.dwarfs,08FrMaLo.exo,11BaAhMe.exo}. The requirements
for fully characterising the methane at the temperatures found in hot
Jupiters are very demanding. A number of groups
\cite{01ScPaxx.CH4,02Scxxxx.CH4,09WaScSh.CH4,13BaWeSu.CH4,13WaCaxx.CH4,13MiChTr.CH4,13ReNiTy.CH4,13ReNiTy.CH4.i}
have been or are attempting to address
this problem. Very recently we have computed a comprehensive line list
for methane \cite{jt564} which is known as 10to10 since it contains just under
10 billion lines. This line list has been demonstrated to give excellent
results in detailed studies of bright T4.5 brown dwarf 2MASS 0559-14
\cite{14YuTeBa.CH4}. The 10to10 line list is designed to model methane
spectra for temperatures up to 1500~K but will require further work
to be complete for higher temperatures and to improve the representation
of the spectrum at short wavelengths.\\

{\bf CO}: HITEMP contains an extensive CO line list constructed using laboratory and sunspot spectra.
These should be sufficient for exoplanetary work. However, there has been
recent work on IR spectrum \cite{10TaVeMi.CO,12VeMiTa.CO,13CoHa.CO} and the
hot VUV spectra \cite{13BrJoCr.CO}. The ExoMol project is planning new line lists
which should both
provide data for all CO isotopologues and increase the range of the line list to all
bound-bound rotation-vibration transitions which is important for studies
of hotter objects such as cool stars.\\

{\bf CO$_{2}$}: HITEMP contains a comprehensive line list for carbon dioxide. However
this line list has been subsequently improved and extended \cite{11TaPe.CO2}. This
new line list should be sufficient for exoplanet studies although
other sources of CO$_2$ data have become available for
both infrared \cite{12HuScTaLe.CO2,13HuFrTa.CO2,13StPiSn.CO2} and
ultraviolet wavelengths \cite{13VeFrBe.CO2}.\\

{\bf FeH}: Experimental partial line lists are available \cite{03DuBaBu.FeH},
further observational data is available from high resolution studies
of M-dwarf star \cite{10HaHiBa.FeH,10WEReSe.FeH} and sunspot \cite{09Faxxxx.FeH} spectra.
These data will be used by ExoMol to make a comprehensive line list.\\

{\bf H$_{2}$}: A comprehensive line list for this molecule covering bound-bound transitions has recently been constructed \cite{12CaKaPa.H2}. A new version
of the
important continuum induced absorption (CIA) has also been released
\cite{12RiGoRo.H2}. These datasets   should be sufficient for exoplanet studies.\\

{\bf H$_{2}$O}: The BT2 line list \cite{jt378} was used to make the original identification
of water in HD189733b \cite{jt400}. Subsequently the BT2 line list was used as the
basis for water in the HITEMP, which also made use of the available laboratory data.
Although work is continuing on improving the representation of water \cite{jt509,jt539,jt550},
the data in HITEMP is both comprehensive and accurate, and should prove the
required spectroscopic data.
\\

{\bf H$_{2}$S}: Hydrogen sulphide is being studied as part of the ExoMol project. A full, high temperature line list, which will be
be sufficient for exoplanetary studies, has just been completed \cite{jth2s}.\\

{\bf H$_{3}^{+}$}: A comprehensive line list for H$_3^+$ was provided some time ago
\cite{jt181} and has already been used for exoplanetary studies \cite{kam07}. A new,
upgraded line list based on an improved theoretical
treatment of this molecule \cite{jt512,jt526} is likely to become available
fairly soon; however the available line list is already sufficient
for exoplanet studies as confirmed by recent experimental tests \cite{jt512,jth3p}.\\

{\bf HCN}: Line lists for hydrogen cyanide were some of the first
calculated using variational nuclear motion calculations
\cite{84ErGuJo.HCN,jt298}. However these line lists are
purely {\it ab initio} and do not give accurate wavelengths.
Harris \ea\ \cite{jt374} improved their line list, which covers both
HCN and HNC, using the then experimental data. A similar
technique was used to create line list for H$^{13}$CN and HN$^{13}$C
\cite{jt447}. Mellau subsequently performed
very extensive experimental studies on both hot HCN and hot HNC
\cite{11Mexxxx.HCN,11Mexxxx.HNC}. A greatly
improved version of the Harris {\it et al} line list has been
constructed using Mellau's energy levels as part of the ExoMol project
\cite{jt570}.
This line list should be sufficient for exoplanet studies.\\

{\bf HDO}:  The VTT line list \cite{jt469} provides a line list for deuterated water
of a quality similar to that of BT2. The accuracy of VTT can be further improved using
experimental data \cite{jt482,14LaVoBo} but is probably already sufficient for planned exoplanet studies.\\

{\bf N$_{2}$O}:  A theoretical, approximate  hot line list in the 4.5 $\mu$m region has been constructed \cite{97RoKhDo.N2O}.  A very comprehensive room temperature line list for N$_2$O is provided by the HITRAN data base \cite{jt557} using laboratory covering all frequency ranges required.
These data should be extended to elevated temperatures.\\

{\bf NH$_{3}$}: Extensive line lists for ammonia are available \cite{jt466,jt500}. The BYTe
line list \cite{jt500} was explicitly designed with the needs of exoplanet spectroscopy in
mind. This line list has already been used for modelling
spectra of brown dwarfs \cite{12BaKexx.dwarfs,jt484} and it should be sufficient for exoplanet studies.\\

{\bf NO$_2$}: A room temperature line list for NO$_2$ is provided by
the HITRAN data base \cite{jt557} and is constructed using laboratory
covering the frequency range required EChO.
These data should be extended to elevated temperatures.\\

{\bf O$_{2}$}: Oxygen spectra have been subject to renewed experimental studied including work at
higher temperatures. This work is captured in the analysis of Yu \ea\ \cite{12YuMiDr.O2}, which
should be sufficient for exoplanet studies.\\

{\bf O$_{3}$}: The spectrum of ozone has been studied systematically over many years in Rheims, France. This
data is captured in the SMPO (Spectroscopy \&\ Molecular Properties of Ozone)  database \cite{SMPO},
which is also accessible as part of the VAMDC project \cite{jt481}. This line list should be sufficient for exoplanet studies.\\

{\bf PH$_{3}$}: Phosphine is being studied as part of the ExoMol project. An initial line list has
recently been released \cite{jt556}.  A full, high temperature line list, which will be
be sufficient for exoplanetary studies, has just been completed \cite{jtph3}.
A more limited, below 3600~\cm\ only, room-temperature, empirical line list
is also available for PH$_3$ \cite{09NiChBu.PH3}.
\\

{\bf SO$_{2}$}:A  line list for SO$_2$ has recently been  computed at NASA Ames \cite{14HuScLe.SO2}.
This line list is for temperatures appropriate to solar system planets (notably Venus). A full  high temperature line list, which will be
sufficient for exoplanetary studies, is being computed as a collaboration between ExoMol and NASA Ames.\\

{\bf TiH}: A partial, empirical line list for TiH is available
\cite{05BuDuBa.TiH}.  Work is in progress as part of the ExoMol
project \cite{jtTiH} producing a comprehensive line list which will
be sufficient for exoplanet studies.\\

{\bf TiO}: A very extensive, hot line list for TiO was constructed by
Schwenke \cite{98Scxxxx.TiO} using a mixture of empirical data and
{\it ab initio} calculations. However, subsequent experimental work on
this system \cite{99RaBeDu.TiO,02KoHaMu.TiO,03NaItDa.TiO} has led to
the suggestion both this and other \cite{98AlPexx.VO} line lists are
incomplete. Despite this criticism, Schwenke's line list has so far
proved sufficient for exoplanetary studies. Given the known importance
of TiO in the spectra of M-dwarf
stars \cite{ahs00}, this problem probably needs to be re-visited.\\

{\bf VO}: Partial experimental line lists are available
\cite{02RaBeDa.VO,05RaBexx.VO}.  Work is already under way in the
ExoMol project which will use
this data to make a comprehensive line list. \\

\section{Other species}

The radicals NO and OH are well-known in the Earth's atmosphere but are not considered in Table~1. Both
are included in HITEMP \cite{jt480} and they
should be included in any comprehensive spectroscopic database.\\

{\bf NO}: The HITEMP  hot line list was constructed using laboratory (lower excitations)  and
extrapolated (high $v, J$) data. NO is characterised by absorption features at 1.1, 1.36, 1.80, 2.7, 5.3 $\mu$m.
The HITEMP line list should be sufficient for exoplanet studies.\\

{\bf OH}:  OH shows absorption features
at 1.0, 1.43, 2.8 $\mu$m. The HITEMP line list should be sufficient for exoplanet studies.\\

Apart from the molecules listed above, ExoMol is undertaking work on
line lists for molecules with the potential to become important for
exoplanetary atmospheric modelling. Some of such molecules are listed
below.\\

{\bf SiO}: ExoMol has released a comprehensive line list which will be sufficient for studies of
the atmospheres of exoplanets and stars \cite{jt563}.\\

{\bf BeH, MgH} and {\bf CaH}: ExoMol has released comprehensive hot line lists for these species  \cite{jt529}.\\

{\bf HCl, NaCl} and {\bf KCl}: are likely to be the main chlorine-bearing species in exoplanets. A new
empirically-derived line list for HCl has been provided by Gordon \ea\ \cite{13LiGoLe.HCl,13LiGoHa.HCl}
which tests show is sufficient for studies at elevated temperatures. New line lists for NaCl and KCl
have been constructed as part of the ExoMol project \cite{jt583}.\\

{\bf SO$_3$}: A room temperature line list has been released by ExoMol \cite{jt554} and a full, high-temperature
list is nearly complete.\\

{\bf H$_2$CO}: A high temperature line list for formaldehyde has just been completed as part of the ExoMol
project \cite{jt584}.\\

{\bf AlH, AlO, C$_2$, C$_3$, CaO, CrH, CS, HNO$_3$, NaH,  NiH, PN, ScH, SH,  SiH}:
Work is  under way in the ExoMol project on these molecules.

There are of course many other species that it may become necessary to
consider including, for example, larger hydrocarbons. Thus, it has has
been suggested that methanol should provide a useful temperature probe
in methane-rich brown dwarfs \cite{02LoFexx.exo}. Partial line lists
are available for methanol \cite{08XuFiLe.CH3OH,12BrSyPe.CH3OH}
although the main focus of these studies is for identifying the highly
complex spectrum of methanol observed in the interstellar medium.

\section{Conclusions}

The EChO mission and other possible attempts to characterise the
atmospheres of exoplanets using spectroscopy place very significant
demands on the laboratory data for spectroscopic interpretation of the
observations. Assembling all the required data requires considerable
effort on the part of laboratory astrophysicists. Much of this data is
already available and can be found on the ExoMol website (www.exomol.com).
Plans for the compiling line lists for the remaining molecules on an
appropriate time scale for the EChO mission are well under way. This
information can be used to construct a mission-specific spectroscopic
database prior to any launch. These data are available
in a variety of formats \cite{jt542} and can be presented in a format
suitable for use in mission-specific software such as EChOSIM
\cite{14WaPaSw.EChO}.

\begin{acknowledgements}
This work was supported by  ERC Advanced Investigator Project 267219 and STFC.


\end{acknowledgements}


\end{document}